# A *Mathematica* package to cope with partially ordered sets

## P. Codara

Dipartimento di Informatica e Comunicazione, Università degli Studi di Milano

---

### Abstract

Mathematica offers, by way of the package *Combinatorics*, many useful functions to work on graphs and ordered structures, but none of these functions was specific enough to meet the needs of our research group. Moreover, the existing functions are not always helpful when one has to work on new concepts.

In this paper we present a package of features developed in *Mathematica* which we consider particularly useful for the study of certain categories of partially ordered sets. Among the features offered, the package includes:

(1) some basic features to treat partially ordered sets;
(2) the ability to enumerate, create, and display monotone and regular partitions of partially ordered sets;
(3) the capability of constructing the lattices of partitions of a poset, and of doing some useful computations on these structures;
(4) the possibility of computing products and coproducts in the category of partially ordered sets and monotone maps;
(5) the possibility of computing products and coproducts in the category of forests (disjoint union of trees) and open maps (cf. [DM06] for the product between forests).

---

## 1. Introduction

In our research we often deal with combinatorial problems, sometimes quite complex, on ordered structures. Recall that a partially ordered set (poset, for short) is a set together with a reflexive, transitive, and antisymmetric binary relation (usually denoted by $\leqslant$) on the elements of the set. Our attention is mainly addressed to certain categories of posets which have strong connections with logic, especially with many-valued (or fuzzy) logic.

In [Cod08] and [Cod09], the author investigates the notion of partition of a poset. In the classical theory, a partition of a set S is a set of nonempty, pairwise disjoint subsets of S, called *blocks*, whose union is S. In the poset case, three different notions of partition are given: monotone, regular, and open. As shown in both works, the "right" notion depends on the category we are dealing with, that is, depends on the kind of posets and maps between them that we are considering. Essentially, a partition (of any kind) of a poset P is a poset whose elements are blocks of a (set) partition of the underlying set of P, endowed with an appropriate partial order. We do not give here the formal definition of the three kinds of partition, and we refer the interested reader to [Cod09].

In [Cod08], it is also shown that the sets of all monotone and regular partitions of a poset can be endowed with lattice structures, i.e., they are poset such that any two elements have a supremum and an infimum.

Many interesting combinatorial questions arise from these two works. For instance: *how many* partitions (of the three kinds) of a poset there are?, *are there* some relations between the three notions, and with the classical notion of partition?, can we give a *simple, intuitive* notion of partition of a poset?, *which properties* the lattices of partitions do have?

The *Mathematica* package presented in this paper was aimed to help answer to these questions, and has been then developed in order to solve more general problems on posets.

In Section 2 we present some basic features of our package *poset.m*. Some of these features are just specialization of some functions avaible via the package *Combinatorica*.

In Section 3 we present some special functions of the package, devoted to the generation, enumeration and investigation of partitions of posets.

In Section 4 we approach the problem of creating the lattice of partitions of a poset, and try to describe how to use some features of *poset.m* to find some properties of the lattice we have obtained.

In Section 5 we present two case studies (taken from [Cod08]) to show how to use the package in two different concrete examples.

Among the features available in the package, we have mentioned the possibility of computing products and coproducts (categorical sum) between posets. In [DM06] the authors present a method to compute coproducts of finitely presented Gödel algebra (particular algebras strictly related to Gödel logic, a many-valued logic). Such method is based on computing a product between forests, i.e posets such that the under set of each element is a totally ordered set (a *chain*). The package *poset.m* include an algorithm to compute this kind of product, besides of course other functions to compute Cartesian products, and coproducts.

In the final Section 6 we show how to use the features just above mentioned.

---

## 2. Basic features

We use the usual command to load the package.

```
<< poset.m
```

First, we want to create a poset. A poset is internally represented as a *Graph* object. The function *Poset* allows to create a poset.

```
? Poset
```

Poset[*relation*] and Poset[*relation*],*v*] generate a partially ordered set, represented
   as a directed graph (see *Combinatorica* manual). *relation* is a set of pairs representing the
   order relation of the poset (not necessarily the entire order relation). *v* is a list of vertices.

```
B2 = Poset[{{"x", "y"}, {"x", "z"}}]
```

◼ Graph:< 5,3,Directed >◼

To display the poset we use the function *Hasse*.

**Hasse[B2]**

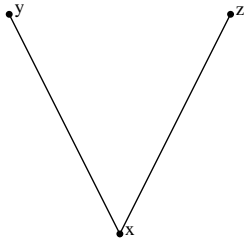

The package provide functions to allow an easy genaration of some common posets.

**C3 = Poset[{{c1, c2}, {c2, c3}}]**

- Graph:< 6,3,Directed >-

**CU3 = Chain[3]**

- Graph:< 6,3,Directed >-

**Hasse[{C3, CU3}]**

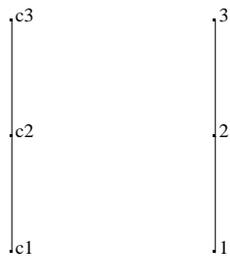

Calling the function *Poset[relation, v]* allows the creation of posets with isolated points. It is also possible to create empty posets.

**F1 = Poset[{{"1", "x"}}, {"0", "1", "x"}]**

- Graph:< 4,3,Directed >-

**Hasse[F1]**

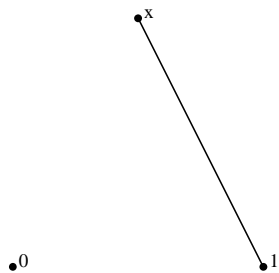

The package provide functions to get different representation of a poset, and to investigate its structure (elements, order relation,...).

**? Relation**

Relation[*p*] gives the list of pairs of the order relation of the partially ordered set *p*. *p* is a *Poset*, or a list of *Posets*.

```
Relation[C3]
```

$\{\{c_1, c_1\}, \{c_1, c_2\}, \{c_1, c_3\}, \{c_2, c_2\}, \{c_2, c_3\}, \{c_3, c_3\}\}$

```
? Covering
```

Covering[*p*] gives the list of pairs of the covering relation of the partially ordered set *p*. *p* is a *Poset*, or a list of *Posets*.

```
Covering[C3]
```

$\{\{c_1, c_2\}, \{c_2, c_3\}\}$

```
PosetElements[C3]
```

$\{c_1, c_2, c_3\}$

---

## 3. Investigating partitions of posets

In this section we show how to work on partitions of a posets using *poset.m*. At the moment, the package allows to create and manage monotone and regular partitions.

```
? PosetPartitions
```

PosetPartitions[*p*] generates the list of all monotone partitions of a poset. Each monotone partition is represented as a *graph*. *p* is a *poset*. PosetPartitions[*p*] outputs a list of *graphs*, and displays the total number of monotone partition of *p*, and the total number of cases analyzed by the function to obtain the monotone partitions.

```
PB2 = PosetPartitions[B2];
```

Analyzed preorders: 16 - Poset Partitions: 7

The function *CreateHasse* allows to display the monotone partitions as posets. In the following Hasse diagrams, the labels of the elements represent blocks of the partitions. They are obtained by concatenating the labels of the elements of each block in the original poset.

```
CreateHasse[PB2, 3]
```

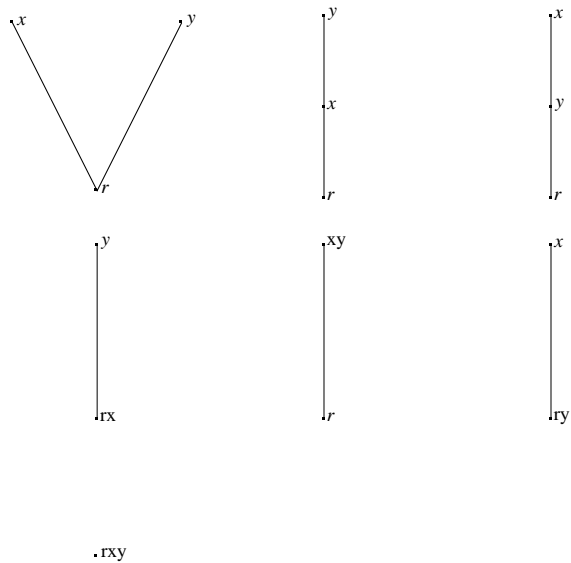

In general, a monotone partitition returned by the function *PosetPartitions* is not a poset, but a preorder (i.e. the binary relation does not have the antisymmetric property). If we apply the function *Hasse* to one of such partitions, we do not obtain an Hasse diagram, but a directed graph. The function *PartitionToPoset* solves this problem, by reducing blocks to single elements and concatenating labels as mentioned above.

**Hasse[PB2[[4]]]**

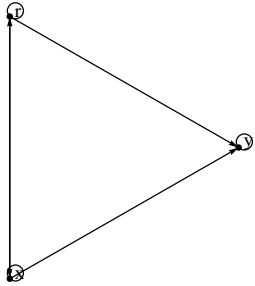

**? PartitionToPoset**

PartitionToPoset[*p*] generates the poset corresponding to a preorder seen as a partition. Whenever two elements *a* and *b* are such that (*a*,*b*) and (*b*,*a*) belong to the order relation of *p*, they are elements of the same block, and they are transformed in a single element of a new poset. *p* is a *graph*, or a list of graphs.

**Hasse[PartitionToPoset[PB2[[4]]]]**

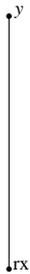

The analogous of *PosetPartition* for generating regular partitions of posets is the function *RegularPartitions*.

**? RegularPartitions**

RegularPartitions[*p*] generates the list of all regular partitions of a poset. Each regular partition is represented as a *graph*. *p* is a *poset*. RegularPartitions[*p*] outputs a list of *graphs*, and displays the total number of regular partition of *p*, and the total number of cases analyzed by the function to obtain the regular partitions.

**RB2 = RegularPartitions[B2];**

Analyzed preorders: 5 – Regular Partitions: 5

```
CreateHasse[RB2, 3]
```

y      z          z          yz

x          xy          x

y

xz          xyz

---

## 4. Investigating the lattices of partitions

As for set partitions, monotone (regular) partitions of a poset can be endowed with a lattice structure. This can be done by ordering the monotone (regular) partitions of the poset by inclusion between the preorder relations representing each partition (cf. [Cod08, 4]). In the following we see how to construct and manage the monotone and regular partition lattices of a poset with the package *poset.m*. Some additional features are also provided. For instance, we show the use of two functions, *Upsets* and *Downsets*, allowing to obtain the upper set (or upset) and the lower set (or downset) of each partition in the lattice. These functions outputs lists of positions, each of which gives the positions of the partitions in the input list. The input list is usually obtained by the functions *PosetPartitions* or *RegularPartitions* described in the previous section.

```
? Upsets
```

Upsets[*plist*] returns the list of the upsets of each element of *plist* in the lattice of partition. *plist* is a list of
monotone or regular partitions, and can be obtained by the functions *RegularPartitions* or *PosetPartitions*.

```
Upsets[RB2]
```

```
{{1, 2, 3, 4, 5}, {2, 5}, {3, 5}, {4, 5}, {5}}
```

```
Downsets[RB2]
```

```
{{1}, {1, 2}, {1, 3}, {1, 4}, {1, 2, 3, 4, 5}}
```

```
? PosetPartitionLattice
```

PosetPartitionLattice[*plist*] returns the lattice structure of a given list of monotone or
regular partitions *plist*. *plist* is usually obtained by using *RegularPartitions* or *PosetPartitions*.

```
LRB2 = PosetPartitionLattice[RB2]
```

- Graph:< 12,5,Directed >-

The labels of the elements in the following Hasse diagram indicate the position of the regular partitions in the list *RB2* (see the previous section for the Hasse diagrams of the elements of *RB2*).

**`Hasse[LRB2]`**

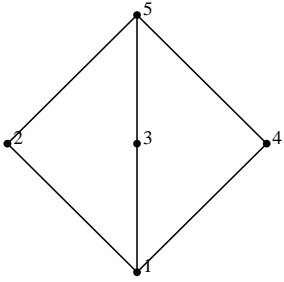

**`Hasse[LPB2 = PosetPartitionLattice[PB2]]`**

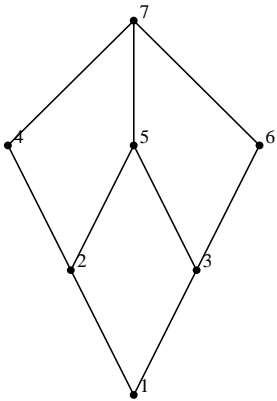

Some functions provided by the package can help in investigating the properties of monotone and regular partitions lattices.

**`? Moebius`**

---

Moebius[*plist*] returns the values of the Moebius

    function computed on each element of a monotone or regular partition lattice.

*plist* is a list of partitions usually obtained by using the functions *PosetPartitions* or *RegularPartitions*.

**`Moebius[PB2]`**

`{1, -1, -1, 0, 1, 0, 0}`

Atoms and coatoms are the elements of the lattice that lie just above the bottom element, or below the top element, respectively.

**`? AtomsPosition`**

---

AtomsPosition[*plist*] returns the list of positions of atoms of the monotone or regular partition lattice given in input.

    *plist* is a list of partitions usually obtained by using the functions *PosetPartitions* or *RegularPartitions*.

**`AtomsPosition[PB2]`**

`{2, 3}`

**`CoatomsPosition[PB2]`**

`{4, 5, 6}`

Some elements of the monotone partition lattice of a poset are isomorphic to linear extensions of the poset.

**Hasse[LinearExtensions[PB2]]**

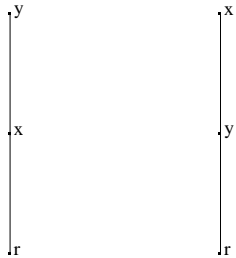

Some lattices (namely, the *ranked* lattices) can be endowed with a rank function $r$ such that $r(x) < r(y)$ whenever $x < y$ and such that whenever $y$ covers $x$, then $r(y) = r(x) + 1$. The value of the rank function for an element of the lattice is called its rank. *Whitney numbers* count the sizes of each level of a ranked lattice, that is, the $n^{\text{th}}$ Whitney number counts the number of elements of a lattice having rank $n$. We assume that the bottom element has rank 1. As shown in [Cod08], the lattice of regular partitions of a poset is always ranked, while the lattice of poset partition, in general, is not.

**? WhitneyNumbers**

WhitneyNumbers[*plist*] returns the Whitney Numbers of a ranked lattice. *plist* is a list of posets forming a ranked lattice.

**WhitneyNumbers[PB2]**

{1, 2, 3, 1}

**? WhitneyLevels**

WhitneyLevels[*plist*] returns the high of the elements of a lattice. *plist* is a list of posets forming a ranked lattice.

**WhitneyLevels[PB2]**

{1, 2, 2, 3, 3, 3, 4}

The latter functions are particularly useful when the graphical output cannot offer any information on the lattice.

**Hasse[P4 = Poset[{{"x", "a"}, {"x", "b"}, {"b", "y"},**
**{"a", "y"}, {"x", "c"}, {"x", "d"}, {"c", "z"}, {"d", "z"}}]]**

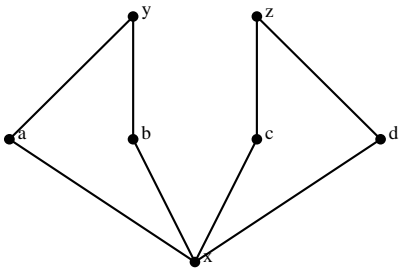

**BigLattice = RegularPartitions[P4];**

Analyzed preorders: 877 - Regular Partitions: 491

**Hasse[PosetPartitionLattice[BigLattice]]**

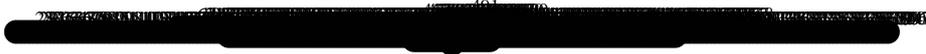

**WhitneyNumbers[BigLattice]**

{1, 19, 107, 208, 131, 24, 1}

**Atoms = AtomsPosition[BigLattice]**

{2, 3, 4, 5, 6, 7, 15, 16, 20, 21, 23, 24, 25, 26, 33, 49, 50, 51, 61}

```
CreateHasse[BigLattice[[Atoms]], 4]
```

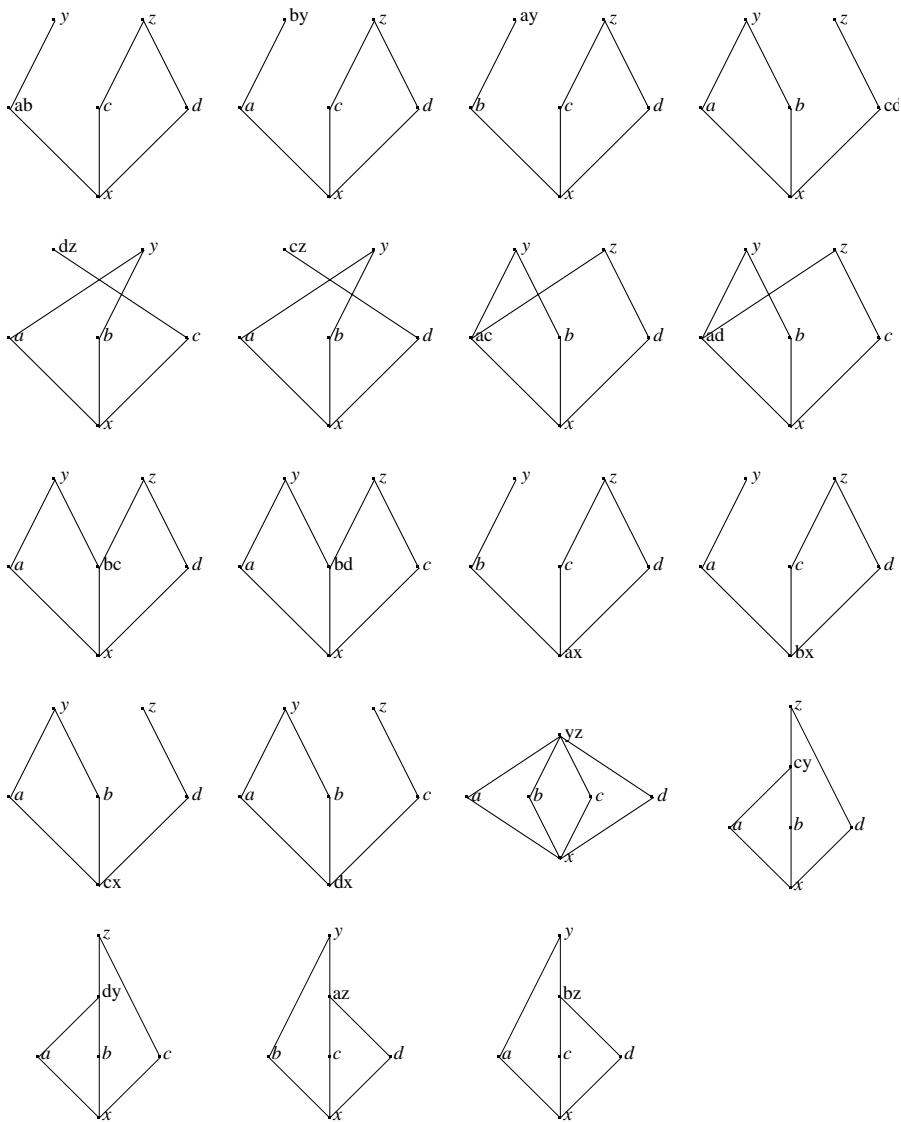

```
Coatoms = CoatomsPosition[BigLattice]
```

{457, 458, 459, 460, 461, 462, 463, 464, 465, 466, 467,
 468, 479, 480, 481, 482, 483, 484, 485, 486, 487, 488, 489, 490}

```
CreateHasse[BigLattice[[Coatoms]], 7]
```

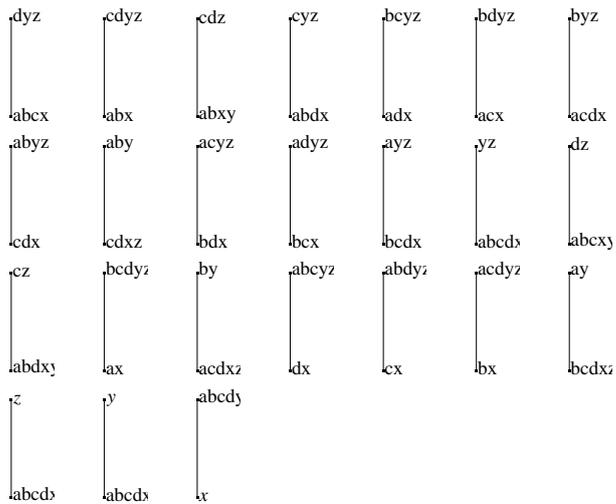

---

# 5. Case studies

## ■ 5.1 Partitions of chains

In [Cod08, 6.2] it is proved that the monotone partition lattice and the regular partition lattice of a chain with $n$ elements are isomorphic, and that they are isomorphic to the Boolean lattice $B_{n-1}$. In this section, we show some small example of this fact.

```
Hasse[{PosetPartitionLattice[PosetPartitions[Chain[2]]],
    PosetPartitionLattice[PosetPartitions[Chain[3]]],
    PosetPartitionLattice[PosetPartitions[Chain[4]]]}]
```

```
Analyzed preorders: 2 - Poset Partitions: 2

Analyzed preorders: 8 - Poset Partitions: 4

Analyzed preorders: 64 - Poset Partitions: 8
```

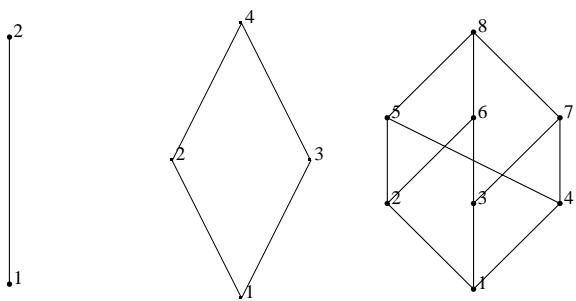

```
Hasse[{PosetPartitionLattice[RegularPartitions[Chain[2]]],
    PosetPartitionLattice[RegularPartitions[Chain[3]]],
    PosetPartitionLattice[RegularPartitions[Chain[4]]]}]
```

```
Analyzed preorders: 2 - Regular Partitions: 2

Analyzed preorders: 5 - Regular Partitions: 4

Analyzed preorders: 15 - Regular Partitions: 8
```

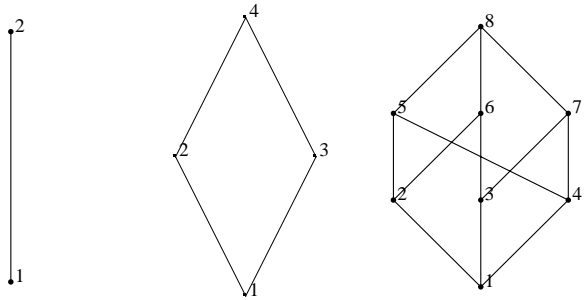

**Hasse[{BooleanAlgebra[1], BooleanAlgebra[2], BooleanAlgebra[3]}]**

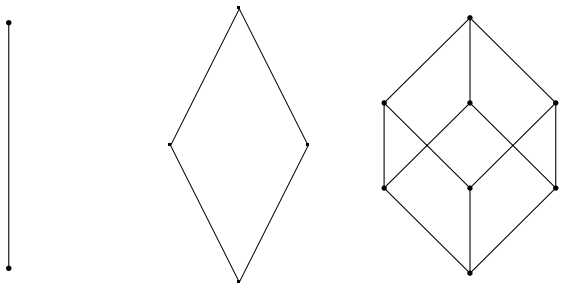

## ■ 5.2 A case of counting

We present in this section an enumerating problem solved in [Cod08, 6.4]. We want to count all regular partitions of a family of posets $M_1$, $M_2$, $M_3$, ..., shown below.

```
M1 = Poset[{{"r", "a"}, {"a", "t"}}];
M2 = Poset[{{"r", "a"}, {"r", "b"}, {"b", "t"}, {"a", "t"}}];
M3 = Poset[{{"r", "a"}, {"r", "b"}, {"r", "c"}, {"c", "t"}, {"b", "t"}, {"a", "t"}}];
M4 = Poset[{{"r", "a"}, {"r", "b"}, {"r", "c"},
    {"r", "d"}, {"d", "t"}, {"c", "t"}, {"b", "t"}, {"a", "t"}}];
M5 = Poset[{{"r", "a"}, {"r", "b"}, {"r", "c"}, {"r", "d"}, {"r", "e"},
    {"e", "t"}, {"d", "t"}, {"c", "t"}, {"b", "t"}, {"a", "t"}}];
```

```
Hasse[{M1, M2, M3, M4, M5}, 3]
```

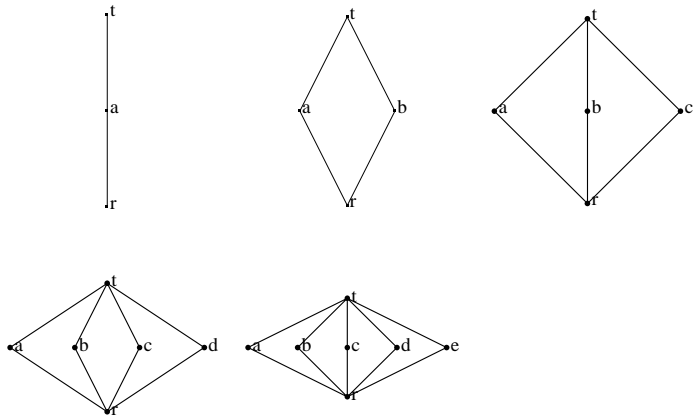

```
RegularPartitions[#] & /@ {M1, M2, M3, M4, M5};
```

```
Analyzed preorders: 5 – Regular Partitions: 4

Analyzed preorders: 15 – Regular Partitions: 11

Analyzed preorders: 52 – Regular Partitions: 38

Analyzed preorders: 203 – Regular Partitions: 152

Analyzed preorders: 877 – Regular Partitions: 675
```

The following formula count the total number of regular partitions of the poset $M_i$ .

$$B_{i+2} - B_{i+1} + 1$$

The number $B_n$ is the $n^{\text{th}}$ Bell number, and it is computed by the *Mathematica* function *BellB[n]*.

```
Table[BellB[n + 2] - BellB[n + 1] + 1, {n, 1, 15}]
```

```
{4, 11, 38, 152, 675, 3264, 17 008, 94 829, 562 596, 3 535 028,
 23 430 841, 163 254 886, 1 192 059 224, 9 097 183 603, 72 384 727 658}
```

## 6. Computing products and coproducts

We present here some examples of computation of products and coproducts of posets. Products and coproducts are different depending on the *category* we are working on. We do not go into details. For the purpose of this work it is sufficient to say that if we consider generic posets as objects (and if maps between posets are order preserving) the coproduct works as a disjoint union, and the product works as a Cartesian product. If, on the other hand, we consider forests as objects (and particular maps between them called *open maps*) the coproduct is a disjoint union and the product can be computed as described in [DM06]. To distinguish the *Mathematica* functions for product and coproducts, their names begin with *Poset* if we are working in the category of posets, and they begin with *Forest* if we are working in the category of forests.

```
F1 = Poset[{{1, 2}}]; F2 = Poset[{{1, 1}, {2, 3}}];
```

**Hasse[F1]**

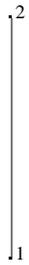

**Hasse[F2]**

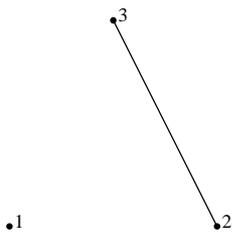

**? PosetSum**

PosetSum[p1,p2,...] computes the sum (coproduct) of the *posets*
p1, p2,... in the category of posets and order preserving maps. The output is a *poset*.

**Hasse[PosetSum[F1, F2]]**

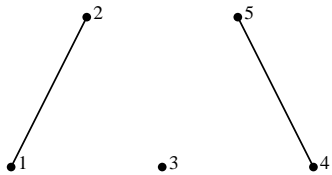

**? ForestSum**

ForestSum[p1,p2,...] computes the sum (coproduct) of the
*forests* p1, p2,... in the category of forests and open maps. The output is a *forest*.

**Hasse[ForestSum[F1, F2]]**

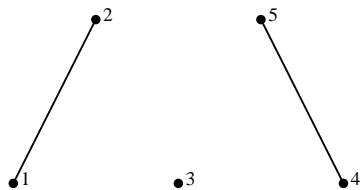

**? PosetProduct**

PosetProduct[p1,p2,...] computes the product of the *posets* p1,
p2,... in the category of posets and order preserving maps. The output is a *poset*.

**`Hasse[PosetProduct[F2, F1]]`**

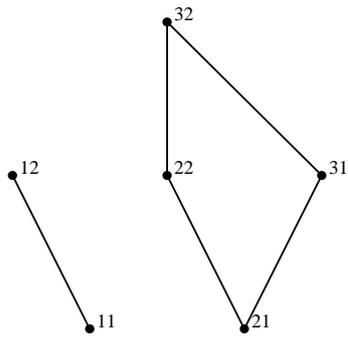

**`? ForestProduct`**

ForestProduct[p1,p2,...] computes the product of the
  *forests* p1, p2,... in the category of forests and open maps. The output is a *forest*.

**`Hasse[ForestProduct[F2, F1]]`**

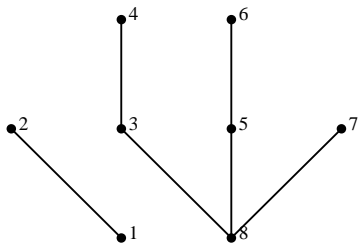

# 7. Acknowledgments

The author would like to thank Silvio Iacobucci who wrote part of the code (related to products and coproducts) during his stage at the Laboratory of Languages and Combinatorics G.-C. Rota (LIN.COM), Università degli Studi di Milano, under the supervision of Prof. O. M. D'Antona.